\documentstyle[12pt]{article}
\def\Qslash{Q\kern-0.15em\raise0.17ex\llap{/}\kern0.15em\relax}
\def\hatQslash{\hat{Q}\kern-0.15em\raise0.17ex\llap{/}
\kern0.15em\relax}
\def\Kslash{K\kern-0.15em\raise0.17ex\llap{/}\kern0.15em\relax}
\def\Pslash{P\kern-0.15em\raise0.17ex\llap{/}\kern0.15em\relax}
\def\Nslash{N\kern-0.15em\raise0.17ex\llap{/}\kern0.15em\relax}
\def\Rslash{R\kern-0.1em\raise0.3ex\llap{/}\kern0.15em\relax}
\def\appendix{%
\section*{Appendix}
 \setcounter{equation}{0}
 \setcounter{section}{1}
 \def\theequation{\mbox{\Alph{section}.\arabic{equation}}}
 }
\begin{document}

\hspace*{65ex} OCU-PHYS-153

\hspace*{65ex} August 1994

\hspace*{65ex} (Revised version)

\hspace*{35ex}

\begin{center}
{\Large {\bf Production of soft photons from the quark-gluon
plasma in hot QCD}} \\

{\Large {\bf --- Screening of mass singularities } }\\

\hspace*{3ex}

\hspace*{3ex}

\hspace*{3ex}

{\large {\sc A. Ni\'{e}gawa}\footnote{
E-mail: h1143@ocugw.cc.osaka-cu.ac.jp}

{\normalsize\em Department of Physics, Osaka City University } \\
  {\normalsize\em Sumiyoshi-ku, Osaka 558, Japan} } \\

\hspace*{2ex}

\hspace*{2ex}

\hspace*{2ex}

{\large {\bf Abstract}} \\
\end{center}

\begin{quotation}

It has been reported that, within the hard-thermal-loop resummation
scheme, the production rate of soft real photons from a hot
quark-gluon plasma exhibits unscreened mass singularities. We show
that still higher-order resummations screen the mass singularities
and obtain the finite soft-photon production rate to leading order
at logarithmic accuracy ${\cal O} (\alpha \alpha_s \ln^2 \alpha_s)$.
\end{quotation}

\newpage
\section{Introduction}
\noindent It has been established by Braaten and Pisarski \cite{bra}
that, in perturbative thermal QCD, the resummations of the
leading-order terms, called hard thermal loops, are necessary. In
thermal massless QCD, we encounter the problem of infrared and mass
or collinear singularities. The hard-thermal-loop (HTL) resummed
propagators soften or screen the infrared singularities, and render
otherwise divergent physical quantities finite, if they are not
sensitive to a further resummation of the corrections of
${\cal O}(g^2 T)$. [For a compact review of infrared and mass
singularities in thermal field theory, we refer to \cite{leb}.]

The calculation of the production rate of soft real photons
$(E={\cal O}(gT))$ from a hot quark-gluon plasma, to leading order
${\cal O} (\alpha \alpha_s \ln \alpha_s)$, within the HTL
resummation scheme has recently been reported \cite{bai,aur} with
the conclusion that the result is divergent, owing to mass
singularities.

The purpose of this paper is to show that resummations of formally
still higher-order contributions screen the above-mentioned mass
singularities and the production rate of ${\cal O}
(\alpha \alpha_s \ln^2 \alpha_s$)) results. In evaluating the
soft-photon production rate, the imaginary-time formalism has been
used in \cite{bai}, while, in \cite{aur}, use has been made of the
retarded/advanced formalism of real-time thermal field theory. For
the purpose of this paper, the ordinary real-time formalism is
convenient, since, in this formalism, identifications of the
physical processes that lead to mass singularities mentioned above
are straightforward \cite{nie2}.

In Sect. 2, we compute the singular part of the production rate of
soft photons to leading order in HTL-resummation scheme of
{\em real-time} thermal field theory. Although the result is known
\cite{bai,aur}, we repeat the calculation in such a manner that the
procedure fits our purpose of further resummation. In Sect. 3,
we carry out further resummations of (formally) higher-order
contributions and show that the mass singularities are screened.
Then, the production rate is evaluated at leading logarithmic
accuracy, i.e., the constant $c$ in $E \, d W / d^{\, 3} p = c
\alpha \alpha_s \ln^2 \alpha_s$ is computed. Sect. 4 is devoted to
discussions and conclusions. Appendix collects some formulae used in
the text.

\section{Leading-order calculation}
After summing over the polarizations of the photon, the production
rate is given by \cite{nie2,kob};
\begin{equation}
E \, \frac{dW}{d^{\, 3} p} = \frac{i}{2 \, (2 \pi)^3} \, g_{\mu \nu}
\, \Pi^{\mu \nu}_{12} (E, {\bf p}) \, .
\label{def}
\end{equation}
In (\ref{def}), $\Pi^{\mu \nu}_{12}$ is the $(1,2)$ component of the
photon polarization tensor in the real-time formalism based on the
time path $C_1 \oplus C_2 \oplus C_3$ in the complex time plane;
$C_1 = - \infty \to + \infty, \, C_2 = + \infty \to - \infty, \,
C_3 = - \infty \to - \infty - i / T$. [The time-path segment $C_3$
does not play \cite{lan,nie3} any explicit role in the present
context.] The fields whose time arguments are lying on $C_1$ and on
$C_2$ are referred, respectively, to as the type-1 and type-2
fields. A vertex of type-1 (type-2) fields is called a type-1
(type-2) vertex. Then $\Pi^{\mu \nu}_{12}$ in (\ref{def}) is the
\lq\lq thermal vacuum polarization between the type-2 photon and the
type-1 photon''. [It is worth mentioning that the \lq\lq Feynman
rules'' of the above-mentioned real-time formalism is equivalent to
the circled diagram rules introduced by Kobes and Semenoff
\cite{kob}, provided that the type-1 (type-2) field is identified
with the field of \lq\lq uncircled'' (\lq\lq circled'') type.]

To leading order, Fig. 1 is the only diagram \cite{bai,aur} that
contributes to $E \, dW /d^{\, 3} p$. In Fig. 1, $p_0 = p = E$ and
2 and 1 at the effective photon-quark vertices stand for the type of
vertices (cf. (\ref{ht-ver}) below). Fig. 1 gives
\begin{eqnarray}
\Pi^{\mu \nu}_{1 2} (P) & = & - \, i \, e_q^2 e^2 N_c \int
\frac{d^{\, 4} K}{(2 \pi)^4} \sum_{i_1, ... , i_4 = 1}^2 tr
   \Big[
       \displaystyle{ \raisebox{0.6ex}{\scriptsize{*}}} \!
          S_{i_1 i_4} (K) \nonumber \\
& & \cdot \left(
      \displaystyle{\raisebox{0.6ex}{\scriptsize{*}}}
      \Gamma^\nu (K, K') \right)^{2}_{i_4 i_3}
      \displaystyle{\raisebox{0.6ex}{\scriptsize{*}}} \!
          S_{i_3 i_2} (K')
      \left( \displaystyle{\raisebox{0.6ex}{\scriptsize{*}}}
      \Gamma^\mu (K', K)
      \right)^{1}_{i_2 i_1}
      \Big] \, ,
\label{pi}
\end{eqnarray}
where $i_1, ... , i_4$ are the thermal indices that specify the
field types. In (\ref{pi}), all the momenta $P, K$ and $K'$ are soft
$(\sim gT)$, so that both photon-quark vertices, $
\displaystyle{\raisebox{0.6ex}{\scriptsize{*}}}
\Gamma^\nu$ and $\displaystyle{\raisebox{0.6ex}{\scriptsize{*}}}
\Gamma^\mu$, as well as both quark propagators, $
\displaystyle{\raisebox{0.6ex}{\scriptsize{*}}} \! S_{i_1 i_4}$ and
$\displaystyle{\raisebox{0.6ex}{\scriptsize{*}}} \! S_{i_3 i_2}$,
are HTL-resummed effective ones (cf. (\ref{ht-ver}) and (\ref{eff1})
- (\ref{eff9})). [Throughout this paper, a capital letter like $P$
denotes the four momentum, $P = (p_0, {\bf p})$, and a lower-case
letter like $p$ denotes the length of the three vector,
$p = |{\bf p}|$. The unit three vector along the direction of, say,
${\bf p}$ is denoted as $\hat{{\bf p}} \equiv {\bf p}/p$. The null
four vector like $\hat{P}$ is defined as $\hat{P} =
(1, \hat{{\bf p}})$].

For our purpose, it is convenient to decompose $g_{\mu \nu}$ in
(\ref{def}) into two parts as
\begin{eqnarray}
& & g_{\mu \nu} = g^{(t)}_{\mu \nu} + g^{(\ell)}_{\mu \nu}
\label{g0} \\
& & g_{\mu \nu}^{(\ell)} = g_{\mu 0} \hat{P}_\nu + g_{\nu 0}
\hat{P}_\mu - \hat{P}_\mu \hat{P}_\nu \, .
\label{g}
\end{eqnarray}
Substituting (\ref{g0}) for $g_{\mu \nu}$ in (\ref{def}), we have,
with an obvious notation,
\begin{equation}
E \, \frac{dW}{d^{\, 3} p} = E \, \frac{dW^{(t)}}{d^{\, 3} p} +
E \, \frac{dW^{(\ell)}}{d^{\, 3} p} \, .
\label{div}
\end{equation}
Now we observe that
$\left( \displaystyle{\raisebox{0.6ex}{\scriptsize{*}}}
\Gamma^\mu \right)^1_{j i}$ in (\ref{pi}) is written as
\begin{eqnarray}
\left( \displaystyle{\raisebox{0.6ex}{\scriptsize{*}}} \Gamma^\mu
\right)^1_{j i} & = & \gamma_\mu \, \delta_{1 j} \, \delta_{1 i} +
\left( \displaystyle{\raisebox{0.6ex}{\scriptsize{*}}}
\tilde{\Gamma}^\mu \right)^1_{j i} \, , \nonumber \\
\left( \displaystyle{\raisebox{0.6ex}{\scriptsize{*}}}
\tilde{\Gamma}^\mu \right)^1_{j i}
& = & \int d \Omega \, \hat{Q}_1^\mu \hatQslash_1 \,
f^1_{j i} (\hat{Q}_1, K', K) \, , \mbox{\hspace{6ex}} (\ell = 1, 2)
\, ,
\label{gamma}
\end{eqnarray}
where $\left( \displaystyle{\raisebox{0.6ex}{\scriptsize{*}}}
\tilde{\Gamma}^\mu \right)^1_{j i}$ is the HTL correction, in terms
of an angular integral, and $Q^\mu_1 = q_1 \hat{Q}_1^\mu$ is the
hard momentum circulating along the HTL. Here, the explicit form of
the function $f^1_{j i}$ in (\ref{gamma}) is not necessary. The
expression for
$\left( \displaystyle{\raisebox{0.6ex}{\scriptsize{*}}} \Gamma^\nu
\right)^2_{j i}$ is given by (\ref{gamma}) with $\gamma^\mu \,
\delta_{1 j} \, \delta_{1 i} \to - \gamma^\nu \, \delta_{2 j} \,
\delta_{2 i}$, $\hat{Q}_1^\mu \to \hat{Q}_2^\nu$, $\hatQslash_1 \to
\hatQslash_2$, and $f^1_{j i} (\hat{Q}_1 , K', K) \to f^2_{j i}
(\hat{Q}_2, K, K')$. It turns out that the production rate
$E \, dW/d^{\, 3} p$ diverges \cite{bai,aur} due to the mass
singularities. As in \cite{bai,aur}, we are only interested in the
divergent parts neglecting all finite contributions.

\hspace*{1ex}

\noindent {\em Computation of $E \, dW^{(t)}/d^{\, 3} p$}

\hspace*{1ex}

\noindent As is seen below, the mass singularities arise
\cite{bai,aur} from the factors $(1 - \sigma \hat{{\bf p}} \cdot
\hat{{\bf q}}_1)^{-1} \, (\sigma = \pm)$ and/or $(1 - \sigma'
\hat{{\bf p}} \cdot \hat{{\bf q}}_2)^{-1} \, (\sigma' = \pm)$, which
come from $\left( \displaystyle{\raisebox{0.6ex}{\scriptsize{*}}}
\tilde{\Gamma}^\mu \right)^1_{i_2 i_1}$  and/or
$\left( \displaystyle{\raisebox{0.6ex}{\scriptsize{*}}}
\tilde{\Gamma}^\nu \right)^2_{i_4 i_3}$ in (\ref{pi}) (cf.
(\ref{gamma})). When integrating over the directions of
$\hat{{\bf q}}_1$ and/or $\hat{{\bf q}}_2$, these factors lead to
divergences at $\hat{{\bf p}} \cdot \hat{{\bf q}}_1 = \sigma$ and/or
at $\hat{{\bf p}} \cdot \hat{{\bf q}}_2 = \sigma'$. Let us see the
numerator factors in the integrand of $E \, dW^{(t)}/d^{\, 3} p$,
which are obtained after taking the trace of Dirac matrices under
the HTL approximation (cf. (\ref{pi}) - (\ref{gamma})).

\begin{itemize}
\item
When one of the photon-quark vertices in Fig. 1 is the HTL
correction and the other is the bare vertex, it can be shown that
the numerator is a linear combination of the terms, each of which
contains the combination $(\hat{{\bf q}}_j \cdot {\bf a})$ $-$
$(\hat{{\bf q}}_j \cdot \hat{{\bf p}}) (\hat{{\bf p}} \cdot
{\bf a})$, $j = 1$ or $2$, with ${\bf a}$ being $\hat{{\bf q}}_j$ or
$\hat{{\bf k}}$ or $\hat{{\bf k}}'$. Then, the numerator vanishes at
$\pm {\bf q}_j \parallel {\bf p}$, meaning that there is no
singularity in the integrand.
\item
When both photon-quark ver\-tices in Fig. 1 are the HTL
cor\-rec\-tions, there emerges the numerator factor
$\hat{{\bf q}}_1 \cdot \hat{{\bf q}}_2 - (\hat{{\bf q}}_1 \cdot
\hat{{\bf p}}) (\hat{{\bf p}} \cdot \hat{{\bf q}}_2)$. For
$\hat{{\bf q}}_1 = \sigma \hat{{\bf p}}$ $(\sigma = \pm)$ {\em and}
$\hat{{\bf q}}_2 \not= \sigma' \hat{{\bf p}}$ $(\sigma' = \pm)$, or
for $\hat{{\bf q}}_2 = \sigma' \hat{{\bf p}}$ {\em and} $\,
\hat{{\bf q}}_1 \not= \sigma \hat{{\bf p}}$, this numerator factor
vanishes. For $\hat{{\bf q}}_1 = \sigma \hat{{\bf p}}$ {\em and}
$\hat{{\bf q}}_2 = \sigma' \hat{{\bf p}}$, the numerator factor
gives $(1 -$ $\sigma \hat{{\bf p}} \cdot \hat{{\bf q}}_1)^{1/2} (1 -
\sigma' \hat{{\bf p}} \cdot \hat{{\bf q}}_2)^{1/2}$, and the
integrations over the directions of $\hat{{\bf q}}_1$ and
$\hat{{\bf q}}_2$ converge.
\end{itemize}
Thus, $E \, d W^{(t)} / d^{\, 3} p$ is free from singularity.

\hspace*{1ex}

\noindent {\em Computation of $E \, dW^{(\ell)} / d^{\, 3} p$}

\hspace*{2ex}

\noindent Let us compute $i g^{(\ell)}_{\mu \nu} \,
\Pi^{\mu \nu}_{1 2}$ with $g^{(\ell)}_{\mu \nu}$ as given in
(\ref{g}). It is not difficult to see that the term with
$\hat{P}_\mu \hat{P}_\nu$ in (\ref{g}) does not yield mass-singular
contribution (see below). Thus we obtain, for the singular
contributions,
\begin{eqnarray}
i \, g^{(\ell)}_{\mu \nu} \, \Pi^{\mu \nu}_{1 2} (P)  & \simeq &
e_q^2 e^2
N_c \left[ g_{\mu 0} \hat{P}_\nu + \hat{P}_\mu g_{\nu 0} \right]
\int \frac{d^{\, 4} K}{(2 \pi)^4}
\sum_{i_1, ... , i_4 = 1}^2 tr  \Big[
\displaystyle{ \raisebox{0.6ex}{\scriptsize{*}}} \! S_{i_1 i_4} (K)
\nonumber \\
& & \cdot
    \left(
      \displaystyle{\raisebox{0.6ex}{\scriptsize{*}}}
      \Gamma^\nu (K, K')
      \right)^{2}_{i_4 i_3}
      \displaystyle{\raisebox{0.6ex}{\scriptsize{*}}} \!
          S_{i_3 i_2} (K')
      \left( \displaystyle{\raisebox{0.6ex}{\scriptsize{*}}}
      \Gamma^\mu (K', K)
      \right)^{1}_{i_2 i_1}
      \Big] \, ,
\label{pil1}
\end{eqnarray}
where the symbol \lq\lq $\simeq$'' is used to denote an
approximation that is valid for keeping the singular contributions.
Using the Ward-Takahashi relations, (\ref{ward}), we can simplify
(\ref{pil1}) as
\begin{eqnarray}
i \, g^{(\ell)}_{\mu \nu} \Pi^{\mu \nu}_{1 2} (P)  & \simeq &
e_q^2 e^2 N_c \frac{1}{p} \int \frac{d^{\, 4} K}{(2 \pi)^4}
\sum_{i = 1}^2
\left\{  tr
\left[
\displaystyle{ \raisebox{0.6ex}{\scriptsize{*}}} \! S_{2 i} (K')
\left( \displaystyle{\raisebox{0.6ex}{\scriptsize{*}}}
      \Gamma^0 (K', K) \right)^{1}_{i 2}
\right.
\right.
\nonumber \\
& & \left.
\mbox{\hspace{24.3ex}} -
      \left( \displaystyle{\raisebox{0.6ex}{\scriptsize{*}}}
      \Gamma^0 (K', K) \right)^{1}_{2 i}
\displaystyle{ \raisebox{0.6ex}{\scriptsize{*}}} \! S_{i 2} (K)
\right] \nonumber \\
& & \left.
+ tr
\left[
      \left( \displaystyle{\raisebox{0.6ex}{\scriptsize{*}}}
      \Gamma^0 (K, K') \right)^{2}_{1 i}
\displaystyle{ \raisebox{0.6ex}{\scriptsize{*}}} \! S_{i 1} (K')
- \displaystyle{ \raisebox{0.6ex}{\scriptsize{*}}} \! S_{1 i} (K)
      \left( \displaystyle{\raisebox{0.6ex}{\scriptsize{*}}}
      \Gamma^0 (K, K') \right)^{2}_{i 1}
\right]
\right\} \, .
\label{pil2}
\end{eqnarray}
We may easily see that the two terms in each set of square brackets
give the same contributions, and the mass-singular contributions
arises from the HTL-corrected parts (eq. (\ref{gamma}))
$\displaystyle{\raisebox{0.6ex}{\scriptsize{*}}}
\tilde{\Gamma}^0$'s of
$\displaystyle{\raisebox{0.6ex}{\scriptsize{*}}} \Gamma^0$'s in
(\ref{pil2}). It is also not difficult to see that the mass
singularities arise \cite{bai,aur} from the terms with
$\left( \displaystyle{\raisebox{0.6ex}{\scriptsize{*}}}
\tilde{\Gamma}^0 \right)^{\ell}_{i j}$ with $i \neq j$.

For the sake of convenience of the calculation in the next section,
we do not use the known simple expressions for
$\displaystyle{\raisebox{0.6ex}{\scriptsize{*}}}
\tilde{\Gamma}^\mu$'s, but rather use their defining expressions
(cf. Fig. 2):
\begin{eqnarray}
\left( \displaystyle{\raisebox{0.6ex}{\scriptsize{*}}}
      \tilde{\Gamma}^\mu (K', K) \right)^{1}_{2 1}
& = & - \, i \, g^2 \,
\frac{N_c^2 - 1}{2 \, N_c} \int \frac{d^{\, 4} Q}{(2 \pi)^4}
\, \gamma_\xi \, S_{21} (Q) \, \gamma^\mu \, S_{11} (Q + P) \,
\gamma_\zeta \, \Delta^{\xi \zeta}_{12} (Q - K') \, , \nonumber \\
\left( \displaystyle{\raisebox{0.6ex}{\scriptsize{*}}}
      \tilde{\Gamma}^\nu (K, K') \right)^{2}_{2 1}
& = & i \, g^2 \,
\frac{N_c^2 - 1}{2 \, N_c} \int \frac{d^{\, 4} Q}{(2 \pi)^4}
\, \gamma_\xi \, S_{22} (Q + P) \, \gamma^\nu \, S_{21} (Q) \,
\gamma_\zeta \, \Delta^{\xi \zeta}_{12} (Q - K') \, , \nonumber \\
\label{ht-ver}
\end{eqnarray}
where $S_{j i}$ and $\Delta^{\xi \zeta}_{1 2}$ are the bare thermal
propagators of a quark and a gluon, respectively (cf. Appendix). In
(\ref{ht-ver}), $Q$ is hard while $P, K$ and $K'$ are soft.
Inserting (\ref{ht-ver}) with $\mu = \nu = 0$ into (\ref{pil2}), we
obtain
\begin{eqnarray}
i \, g^{(\ell)}_{\mu \nu} \, \Pi^{\mu \nu}_{1 2} (P) & \simeq &
4 \, i \, g^2 \, e_q^2 \, e^2 \, N_c \, \frac{N_c^2 - 1}{2 \, N_c}
\, \frac{1}{E} \int \frac{d^{\, 4} K}{(2 \pi)^4} \int
\frac{d^{\, 4} Q}{(2 \pi)^4} \, \Delta_{1 2} (Q- K') \,
\tilde{S}_{2 1} (Q) \nonumber \\
& & \cdot \sum_{\sigma, \sigma' = \pm} tr \left\{
\hat{\Kslash}_\sigma \Qslash \, \gamma^0 \, \Qslash_{\sigma'}
\right\}
\displaystyle{ \raisebox{0.6ex}{\scriptsize{*}}} \!
\tilde{S}^{(\sigma)}_{1 2} (K) \left\{ \tilde{S}_{1 1} (Q + P) -
\tilde{S}_{2 2} (Q + P) \right\} \, , \nonumber \\
\label{pil3}
\end{eqnarray}
where $\hat{K}_\sigma = (1, \sigma \hat{{\bf k}})$. In obtaining
(\ref{pil3}), use has been made of (\ref{eff1}), (\ref{ferb1}) and
(\ref{bosb1}), and, as usual, $Q + P \simeq Q - K' \simeq Q$ in the
factor under the \lq\lq $tr$''. Substituting (\ref{eff4}),
(\ref{eff10}), (\ref{ferb3}) and (\ref{bosb2}) into (\ref{pil3}), we
have
\begin{eqnarray}
i \, g^{(\ell)}_{\mu \nu} \, \Pi^{\mu \nu}_{1 2} (P) & \simeq &
32 \, \pi^3 \, g^2 \, e_q^2 \, e^2 \, N_c \,
\frac{N_c^2 - 1}{2 \, N_c} \, \frac{1}{E} \int
\frac{d^{\, 4} K}{(2 \pi)^4} \, n_F (k_0) \int
\frac{d^{\, 4} Q}{(2 \pi)^4} \, n_B (q_0) \, n_F (- q_0) \,
\nonumber \\
& & \cdot \delta \left( Q^2 \right) \delta \left( (Q - K')^2 \right)
\, \sum_{\sigma, \tau = \pm} \left[ \left( \hat{K}_\sigma
\cdot Q \right) + q^0 \left( \hat{K}_\sigma \cdot \hat{Q}_{\tau}
\right) - \left( Q \cdot \hat{Q}_{\tau} \right) \right] \nonumber \\
& & \cdot \mbox{$\large{\rho}_\sigma$} (K)
\left[ \frac{1}{q_0 (1 + i \epsilon) + p_0 -
\tau |{\bf q} + {\bf p}|} \, + \mbox{ c.c.} \right]  \, ,
\nonumber \\
\label{pil4}
\end{eqnarray}
where $\hat{Q}_\tau = (1, \tau \hat{{\bf q}})$ and the HTL
approximation $\epsilon (q_0 - k'_0) n_B (q_0 - k'_0) \simeq
\epsilon (q_0) n_B (q_0)$ has been made. In (\ref{pil4}),
$\large{\rho}_{\pm}$ (eq. (\ref{eff7})) is the absorptive part of
the effective thermal quark propagator. Due to the factor
$\delta (Q^2)$ in (\ref{pil4}), we have two contributions, the one
with $q_0 = q$ and the other with $q_0 = - q$. The former (latter)
contribution is singular when $\tau = +$ $(-)$;
\begin{equation}
q_0 + p_0 - \tau |{\bf q} + {\bf p}| \simeq E \,
(1 - \tau \, \hat{{\bf p}} \cdot \hat{{\bf q}}) =
E \, (1 - \epsilon (q_0) \, \hat{{\bf p}} \cdot \hat{{\bf q}})\, .
\label{approx}
\end{equation}

When $\hat{P}_\mu \hat{P}_\mu$ term in $g_{\mu \nu}^{(\ell)}$
(eq. (\ref{g})) is substituted for $g_{\mu \nu}$ in $i g_{\mu \nu}
\Pi^{\mu \nu}_{1 2}$, we obtain the Dirac trace $tr \left\{
\Kslash_\sigma \Qslash \hat{\Pslash} \Qslash \right\} = 4\left\{ 2
(K_\sigma \cdot Q) (\hat{P} \cdot Q) - (K_\sigma \cdot \hat{P}) \,
Q^2 \right\}$. The $Q^2$ term gives vanishing contribution due to
$\delta ( Q^2 )$ (cf. (\ref{pil4})), and $\hat{P} \cdot Q = q_0 -
\hat{{\bf p}} \cdot {\bf q} = q_0 ( 1 - \epsilon (q_0) \,
\hat{{\bf p}} \cdot \hat{{\bf q}})$ eliminates the singularity (cf.
(\ref{approx})) of the factors in the square brackets in
(\ref{pil4}) and no mass singularity arises.

Carrying out the integration over $q_0$, we obtain
\begin{eqnarray}
i \, g^{(\ell)}_{\mu \nu} \Pi^{\mu \nu}_{1 2} (P) & \simeq &
g^2 \, e_q^2 \, e^2 \, N_c \, \frac{N_c^2 - 1}{2 \, N_c} \,
\frac{1}{E^2} \, \frac{1}{\pi} \int \frac{d^{\, 4} K}{(2 \pi)^3} \,
\sum_{\sigma = \pm} \mbox{$\large{\rho}_\sigma$} (K)
\nonumber \\
& & \cdot \int q^2 d q \int d(\hat{{\bf p}} \cdot \hat{{\bf q}})
\sum_{\tau = \pm} \tau \left( 1 - \sigma \tau \hat{{\bf q}} \cdot
\hat{{\bf k}} \right) \left\{ \frac{1}{1 - \tau \hat{{\bf p}} \cdot
\hat{{\bf q}} + i \epsilon} + \mbox{c.c.} \right\}
\nonumber \\
& & \cdot \Bigg[ \left\{ \theta
(- \tau) + n_B (q) \right\} \left\{ \theta ( \tau) - n_F (q)
\right\} \nonumber \\
& & \cdot \delta \left(
(\tau q + p_0 - k_0)^2 - ({\bf q} + {\bf p} - {\bf k})^2
\right) \Bigg] \, ,
\label{pil5}
\end{eqnarray}
where we have set $n_F (k_0) = \frac{1}{2}$. This is because, in the
HTL approximations, $e^{k_0 / T} = 1 + {\cal O} (g)$.

In (\ref{pil5}), the integration over $\hat{{\bf p}} \cdot
\hat{{\bf q}}$ diverges at $\hat{{\bf q}} = \hat{{\bf p}}$, which is
nothing but the mass singularity. Using the HTL approximation
$p << q$ in (\ref{pil5}), and carrying out the integrations over $q$
and over $\hat{{\bf p}} \cdot \hat{{\bf q}}$ with the small quark
mass ($\mu$) regulator, we have
\begin{equation}
E \, \frac{dW}{d^{\, 3} p} \simeq
\frac{1}{2 \pi} \, e_q^2 \, \alpha \, N_c \, \frac{m_f^2}{E} \,
\ln \left( \frac{E T}{\mu^2} \right) \int
\frac{d^{\, 4} K}{(2 \pi)^3} \, \delta (P \cdot K)
\sum_{\sigma = \pm} (1 - \sigma \hat{{\bf p}} \cdot \hat{{\bf k}})
\, \mbox{$\large{\rho_\sigma}$} (K) \, ,
\label{w1}
\end{equation}
where $m_f$ (eq. (\ref{eff9})) is the thermal mass of a quark. The
expression (\ref{w1}) is valid as far as the singular contribution
is concerned.

We encounter the same integral as in (\ref{w1}) in the hard-photon
production case \cite{bai1}. Here we recall that $K$ is soft $\sim
{\cal O} (g T)$. Then, the upper limit $k^*$ of the integration over
$k$ is in the range, $g T << k^* << T$. (In \cite{bai}, $k^*$ is
chosen as $k^* = T$.) Referring to \cite{bai1}, we
have
\begin{equation}
E \, \frac{dW}{d^{\, 3} p} \simeq
\frac{e_q^2 \, \alpha \, \alpha_s}{4 \pi^2} \, T^2
\left( \frac{m_f}{E} \right)^2
\ln \frac{E T}{\mu^2} \left[ \ln \left( \frac{k^*}{m_f}
\right)^2 - 1.31  \right] \, .
\label{w2}
\end{equation}
Thus, we have reproduced the result reported in \cite{bai,aur}.

\section{Further resummation}

\noindent We study still higher order corrections. In the course of
calculations in the previous section, we need not take care of the
gauge invariance, since HTL corrections are gauge invariant
\cite{bra}. Let us consider adding additional gluon lines to Fig. 1.
In a covariant gauge, the gauge-parameter dependent part of a gluon
propagator $\Delta^{\mu \nu}_{j i} (K)$ carries a factor
$K_\mu K_\nu$. Then, we see \cite{bjo} that, upon summation over all
possible positions on the quark lines, where the gluon can be
inserted, the gauge-parameter dependent parts cancel, thanks to the
trivial identity $(\Qslash + \Kslash)^{-1} \Kslash (\Qslash)^{-1} =
(\Qslash )^{-1} - ( \Qslash + \Kslash )^{-1}$. Thus, we can use the
Feynman gauge.

Here is an observation of crucial importance: {\em The insertions of
gluon lines to the \lq\lq parts'' of Fig. 1, which do not show up
mass singularities, yield higher-order contributions,} i.e., the
contributions which are at least $g$ times the contributions under
consideration. Then, the only part that should be considered to the
order in consideration, ${\cal O} ( \alpha \alpha_s)$ up to
logarithmic factors, is (\ref{pil3}). The same reasoning as above
tells us that only the gluon-line insertions to the quark
propagators with momentum $Q + P$ are the only corrections that we
should take into account, the quark propagators which originate the
mass singularity in (\ref{w2}). This amounts to replacing
$S_{11} - S_{22}$ (cf. \ref{pil3}) with the one-loop
self-energy-part resummed one (see Appendix),
\begin{eqnarray}
& & \displaystyle{ \raisebox{1.1ex}{\scriptsize{$\diamond$}}}
\mbox{\hspace{-0.33ex}}
S_{11} (R) -
\displaystyle{ \raisebox{1.1ex}{\scriptsize{$\diamond$}}}
\mbox{\hspace{-0.33ex}}
S_{22} (R) = \sum_{\tau = \pm}
\Rslash_{\tau} \left\{
\displaystyle{ \raisebox{1.1ex}{\scriptsize{$\diamond$}}}
\mbox{\hspace{-0.33ex}}
\tilde{S}^{(\tau)}_{1 1} (R) -
\displaystyle{ \raisebox{1.1ex}{\scriptsize{$\diamond$}}}
\mbox{\hspace{-0.33ex}}
\tilde{S}^{(\tau)}_{2 2} (R) \right\} \, ,
\mbox{\hspace{4ex}} R = Q + P
\, , \label{S-star1} \\
& & \displaystyle{ \raisebox{1.1ex}{\scriptsize{$\diamond$}}}
\mbox{\hspace{-0.33ex}}
\tilde{S}^{(\tau)}_{11} (R) -
\displaystyle{ \raisebox{1.1ex}{\scriptsize{$\diamond$}}}
\mbox{\hspace{-0.33ex}}
\tilde{S}^{(\tau)}_{22} (R) =
- \, \frac{1}{2} \left[ \frac{1}{D_{\tau} (R) + (r_0 - \tau r)
\tilde{a} + R^2 \, \tilde{b}} \, + \, \mbox{c.c.} \right] \, ,
\label{S-star2}
\end{eqnarray}
where use has been made of (\ref{eff3}) with (\ref{hardeff}). In
(\ref{S-star2}), $D_\tau$ is as in (\ref{eff5}) and $r_0 \, (= q_0
+ p_0)$ should read $r_0 (1 + i \epsilon)$. Explicit expressions for
$\tilde{a} = \tilde{a} (r_0, r)$ and $\tilde{b} = \tilde{b}
(r_0, r)$ are given in (\ref{a-tilde}) - (\ref{int}). We first note
that the delta-function contributions of quasiparticle modes coming
from two terms in (\ref{S-star2}) cancel, since they are pure
imaginary. From (\ref{a-tilde}) - (\ref{int}), we see that
$\tilde{a} (r_0, r)$ and $\tilde{b} (r_0, r)$ have imaginary parts
in the whole range of $r_0$. We also see from (\ref{approx}) that
$r_0 - \tau r = q_0 + p_0 - \tau |{\bf q} + {\bf p}|$ in
(\ref{S-star2}) is non negative. Then, using (\ref{eff5}), we obtain
\begin{equation}
\displaystyle{ \raisebox{1.1ex}{\scriptsize{$\diamond$}}}
\mbox{\hspace{-0.33ex}}
\tilde{S}^{(\tau)}_{11} (R) -
\displaystyle{ \raisebox{1.1ex}{\scriptsize{$\diamond$}}}
\mbox{\hspace{-0.33ex}}
\tilde{S}^{(\tau)}_{22} (R) = \frac{ (r_0 - \tau r)
( 1 + {\cal F}_\tau) - \tau \frac{m_f^2}{r} }{ \left[
(r_0 - \tau r) (1 + {\cal F}_\tau)  - \tau \frac{m_f^2}{r}
\right]^2
+ (r_0 - \tau r)^2 \, {\cal G}_\tau^2 } \, ,
\label{S-star3}
\end{equation}
where
\begin{eqnarray}
{\cal F}_\tau & = & \tau \, \frac{m_f^2}{2 r^2} \, \ln
\frac{r_0 + r}{r_0 - r} - Re \, \tilde{a} - (r_0 + \tau \, r) Re \,
\tilde{b} \, , \nonumber \\
{\cal G}_\tau & = & Im \, \tilde{a} +
(r_0 + \tau \, r) Im \, \tilde{b} \, .
\label{S-star4}
\end{eqnarray}

The factor $\tilde{S}_{1 1} (Q + P) - \tilde{S}_{2 2} (Q + P)$ in
(\ref{pil3}) is replaced by (\ref{S-star3}), which means that the
quantity in the square brackets in (\ref{pil4}) is replaced by two
times (\ref{S-star3}). As discussed above after (\ref{pil4}), the
mass-singular contributions have arisen from the region $r_0 -
\tau r \simeq E \, (1 - \tau \hat{{\bf q}} \cdot \hat{{\bf p}} )
\simeq 0$ (cf. (\ref{approx})). Then, to obtain the leading
contribution from (\ref{pil4}), with the above-mentioned
substitution being made, it is sufficient to carry out the
$\hat{{\bf p}} \cdot \hat{{\bf q}}$ integration over the small
region $\hat{{\bf p}} \cdot \hat{{\bf q}} \simeq \tau$. Setting
$\hat{{\bf p}} \cdot \hat{{\bf q}} = \tau$ in all the factors except
in the factor (\ref{S-star3}), we obtain
\begin{eqnarray}
\int_{\hat{{\bf p}} \cdot \hat{{\bf q}} \simeq 0}
d (\hat{{\bf p}} \cdot \hat{{\bf q}}) \, \mbox{[eq.
(\ref{S-star3})]}
& \simeq & \frac{1}{E} \, \ln \left( \frac{q E}{m_f^2} \right)
\nonumber \\
& \simeq & \frac{1}{2 \, E} \, \ln \left( \frac{1}{\alpha_s} \right)
\, , \label{finite}
\end{eqnarray}
where use has been made of the fact that ${\cal F}_\tau$ and
${\cal G}_\tau$ are of ${\cal O} (g^2)$ (cf. (\ref{S-star4}),
(\ref{a-tilde}) - (\ref{int})), and ${\cal F}_\tau$ and
${\cal G}_\tau$ are ignored on the r.h.s. of (\ref{finite}). The
factor $m_f^2$ in (\ref{finite}) or in (\ref{S-star3}) comes from
the hard loop momentum region in the (momentum-)integral
representation of $\tilde{\Sigma}_F$ in (\ref{sigma}).

Thus, we finally obtain
\begin{eqnarray}
& & E \, \frac{d W}{d^{\, 3} p} \simeq
\frac{e_q^2 \, \alpha \, \alpha_s}{ 8 \, \pi^2} T^2
\left( \frac{m_f}{E} \right)^2 \ln \left( \frac{1}{\alpha_s} \right)
\, \ln \left\{ \frac{c'}{\alpha_s} \left( \frac{k^*}{T} \right)^2
\right\} \, , \label{final} \\
& & c' \simeq \frac{4 \, N_c}{\pi (N_c^2 - 1)} \, e^{-1.31} \simeq
0.129 \, , \mbox{\hspace{5ex}} (N_c = 3) \, . \nonumber
\end{eqnarray}
As is obvious from our derivation, the result (\ref{final}) is valid
at logarithmic accuracy. In order to obtain a full contribution of
${\cal O} (\alpha \alpha_s)$, we need to evaluate all the
contributions, which we ignored in various stages.

\section{Discussions and conclusions}

\noindent Within the HTL resummation scheme, the soft-photon
production rate (\ref{w2}) diverges \cite{bai,aur}. This divergence
arises from mass singularities.

For any given diagram in real-time formalism, general rules of
identifying the physical processes are available \cite{nie2}.
According to the rules, we can identify the physical processes
leading to mass singularities: For $q_0 > 0$ (cf. Fig. 2), an
example of such diagrams is depicted in Fig. 3 with the final-state
cut line $C_1$. In Fig. 3, the left part of the final-state cut line
represents the $S$-matrix element in {\em vacuum theory}, while the
right part represents the $S^*$-matrix element. The group of
particles on top of Fig. 3 stands for spectator particles which are
constituents of the quark-gluon plasma. When ${\bf q} \parallel
{\bf p}$, the exchanged quark, $q (Q + P)$, is on the mass shell,
which causes the mass singularity.

Here it is worth mentioning the following issue. In vacuum theory,
it has long been known \cite{kin} that, when mass singularities
arise in a rate of some reaction, the states that degenerate with
the final state participate in the intermediate states. For such a
case, the general theorem \cite{kin} of cancelling mass
singularities states that, when we sum up a set of reaction rates,
whose initial and final states as a whole include all the degenerate
sets, the cancellations of mass singularities occur and finite
reaction rate results. Physical interpretation of the theorem is
fully discussed in \cite{kin}. It has been proved \cite{nie} that
this theorem also applies to the case of thermal field theory.

We see from Fig. 3 with the final-state cut line $C_1$ that, when
${\bf q} \parallel {\bf p}$, the single-quark state $q (Q + P)$ in
the intermediate state degenerate with the two-particle state, $q (Q)
+ \gamma (P)$, in the final state. Following the theorem stated
above, we consider the processes whose final state is $q (Q + P)$.
Fig. 3 with the final-state cut line $C_2$ is an example of such
processes. We can see \cite{nie2} that such processes are
represented by
\begin{equation}
\int_\Delta \frac{d^{\, 4} P}{(2 \pi)^4} \sum_{\ell = 1}^2 \left[
\Delta_{\ell \ell}^{\mu \nu} \left\{ \Pi_{\ell \ell} (P)
\right\}_{\mu \nu} \right] \, ,
\label{vir}
\end{equation}
where $\Delta$ is the small region where $P^2 \simeq 0$. Although,
presenting the elaborate formulae is not the purpose of this paper,
we can confirm the above-mentioned theorem in the case of the
soft-photon production rate: Within the HTL resummation scheme, the
cancellation of mass singularities occurs between $E \, d W /
d^{\, 3} p$ in (\ref{w2}) and the rate extracted from (\ref{vir}),
and their sum is free from singularity. In the case of, e.g., $q_0
> 0$ (Fig. 3), the former represents the production rate of the
two-particle state $q (Q) + \gamma (P)$, while the latter represents
the production of $q (Q + P)$.

In conclusion, an explicit computation in this paper of the
soft-photon production rate demonstrates that mass singularities due
to the exchanges of (massless) {\em hard} quarks are shielded by
formally higher-order contributions, in a hot quark-gluon plasma.
This screening mechanism allows us to compute the soft-photon
production rate at logarithmic accuracy (cf. (\ref{final})): The
appearance of the second logarithm $\ln \left\{ \frac{c'}{\alpha_s}
\left( \frac{k^*}{T} \right)^2 \right\}$ in (\ref{final}) reflects
the HTL resummed {\em soft} quark propagators
($\displaystyle{ \raisebox{0.6ex}{\scriptsize{*}}} \!
\tilde{S}_{i_1 i_4} (K)$ and
$\displaystyle{ \raisebox{0.6ex}{\scriptsize{*}}} \!
\tilde{S}_{i_3 i_2} (K')$ in (\ref{pil1})), while the first
logarithm $\ln \left( \frac{1}{\alpha_s} \right)$
arises from the resummed {\em hard} quark propagators,
$\displaystyle{ \raisebox{1.1ex}{\scriptsize{$\diamond$}}}
\mbox{\hspace{-0.33ex}} S_{11} (Q + P)$ and
$\displaystyle{ \raisebox{1.1ex}{\scriptsize{$\diamond$}}}
\mbox{\hspace{-0.33ex}} S_{22} (Q + P)$ (cf.
(\ref{S-star1})).

Finally, the following observation is in order. For shielding quark
mass singularities, only the quark propagator with momentum $Q + P$
(cf. Figs. 2 and 3) and the two onshell lines with momenta $P$ and
$Q$ participate, and \lq\lq other parts'' of the diagram are not
directly relevant. Then, the mechanism of shielding quark mass
singularities works universally, being independent of reactions.

\hspace*{1ex}

\noindent {\em Acknowledgement.} I sincerely thank R. Baier for
useful discussions. Were it not for his patient criticism on the
earlier version, this paper would not appear. Thanks are also due
to T. Yoshida for instructing me in the technical aspect of the
\lq\lq quark detection''.

\hspace*{1ex}

\appendix

\noindent Here we display various expressions and useful formulae.

\hspace*{1ex}

\noindent $\bullet$ {\em Effective thermal propagator of a quark
with soft momentum}

\begin{equation}
\mbox{\hspace{-0.8ex}}
\displaystyle{\raisebox{0.6ex}{\scriptsize{*}}} \! S_{j i} (K)
= \sum_{\sigma = \pm}
\hat{\Kslash}_\sigma
\displaystyle{\raisebox{0.6ex}{\scriptsize{*}}} \!
\tilde{S}_{j i}^{(\sigma)} (K)\, , \mbox{\hspace{6ex}}
(j, i = 1, 2) \, ,
\label{eff1}
\end{equation}
\vspace{-2ex}
where
\vspace{-1ex}
\begin{eqnarray}
& & \hat{K}_\sigma  =
(1, \, \sigma \hat{{\bf k}}) \, ,
\nonumber \\
& & \displaystyle{\raisebox{0.6ex}{\scriptsize{*}}} \!
\tilde{S}_{11}^{(\sigma)} (K)
=  - \left(
\displaystyle{\raisebox{0.6ex}{\scriptsize{*}}} \!
\tilde{S}_{22}^{(\sigma)} (K) \right)^* \nonumber \\
& & \mbox{\hspace{8.2ex}} = - \, \frac{1}{2 D_\sigma (K)} \, + \,
i \pi \, \epsilon (k_0) n_F (|k_0|)
\mbox{$\large{\rho}_\sigma$} (K) \, ,  \label{eff3} \\
& & \displaystyle{\raisebox{0.6ex}{\scriptsize{*}}} \!
\tilde{S}_{12 \, (21)}^{(\sigma)} (K) = \pm i \pi \,
n_F (\pm k_0) \, \mbox{$\large{\rho}_\sigma$} (K) \, ,
\label{eff4}
\end{eqnarray}
\vspace{-2ex}
with
\vspace{-0.5ex}
\begin{eqnarray}
& & D_\sigma (K) = - k_0 + \sigma k + \frac{m_f^2}{2 k}
\left[ \left( 1 - \sigma \frac{k_0}{k} \right) \, \ln
\frac{k_0 + k}{k_0 - k}  \, + \, 2 \sigma \, \right] \, ,
\label{eff5} \\
& & \mbox{$\large{\rho}_\sigma$} (K) =
\frac{\epsilon (k_0)}{2 \pi i} \left[ \,
\frac{1}{D_\sigma (k_0 ( 1 + i 0^+) , k)} \, - \,
\frac{1}{D_\sigma (k_0 ( 1- i 0^+), k)} \, \right] \, ,
\label{eff7} \\
& & n_F (x) = \frac{1}{e^{x / T} - 1} \, \, , \nonumber  \\
& & m_f^2 = \frac{\pi \alpha_s}{2} \frac{N_c^2 - 1}{2 N_c} \,
T^2 \, . \label{eff9}
\end{eqnarray}
In (\ref{eff5}), $k_0$ should read $k_0 (1 + i \epsilon)$.
$D_\sigma (K)$ in (\ref{eff5}) is first calculated in \cite{wel}.

\hspace*{1ex}

\noindent $\bullet$ {\em Bare thermal propagator of a quark}

\noindent The bare thermal propagators $S_{j i} (K) \, \,
(j, i = 1, 2)$ are obtained from (\ref{eff1}) - (\ref{eff9}) with
$m_f = 0$ and $\large{\rho}_\sigma$ $\! (K) = \delta
(k_0 - \sigma k)$: In the text, there appears the combination,
\begin{equation}
\tilde{S}^{(\sigma)}_{1 1} (K) - \tilde{S}^{(\sigma)}_{2 2} (K) =
\frac{1}{2} \left[ \frac{1}{k_0 ( 1 + i \epsilon) - \sigma k} +
\mbox{c.c.} \right] \, .
\label{eff10}
\end{equation}
For the purpose of practical use, it is convenient to use the
\lq\lq one-term'' forms;
\begin{eqnarray}
& & S_{j \ell} (K) =
\Kslash
\, \tilde{S}_{j \ell} (K)
\, , \label{ferb1} \\
& & \tilde{S}_{11} (K) = - \left\{ \tilde{S}_{22} (K) \right\}^{*}
\, , \nonumber \\
& & \mbox{\hspace{7ex}} = \frac{1}{K^2 + i \epsilon} \, + \, 2
\pi i \,
n_F (|k_0|) \, \delta (K^2) \, , \nonumber \\
& & \tilde{S}_{12 \, (21)} (K) = 2 \pi i \, \epsilon (\pm k_0)
\, n_F (\pm  k_0) \, \delta (K^2) \, .
\label{ferb3}
\end{eqnarray}

\hspace*{1ex}

\noindent $\bullet$ {\em Bare thermal gluon propagator}

\begin{eqnarray}
& & \Delta_{12}^{\mu \nu} (Q) = - g^{\mu \nu}
\, \Delta_{12} (Q) \, , \label{bosb1} \\
& & \Delta_{12} (Q) =  - \, 2 \pi i \, \epsilon (q_0) \, n_B (q_0)
\, \delta (Q^2) \, ,
\label{bosb2} \\
& & n_B (x) = \frac{1}{e^{x / T} - 1} \, . \nonumber
\end{eqnarray}

\hspace*{1ex}

\noindent $\bullet$ {\em HTL-corrected Ward-Takahashi relations}
\begin{eqnarray}
(K - K')_\mu \sum_{i_2, i_1 = 1}^2
      \displaystyle{\raisebox{0.6ex}{\scriptsize{*}}} \!
          S_{j i_2} (K')
      \left( \displaystyle{\raisebox{0.6ex}{\scriptsize{*}}}
      \Gamma^\mu (K', K) \right)^\ell_{i_2 i_1}
      \displaystyle{\raisebox{0.6ex}{\scriptsize{*}}} \!
          S_{i_1 i} (K)
  =   \delta_{\ell i}
      \displaystyle{\raisebox{0.6ex}{\scriptsize{*}}} \!
          S_{j i} (K')
       - \delta_{\ell j}
       \displaystyle{\raisebox{0.6ex}{\scriptsize{*}}} \!
          S_{j i} (K) \, , \nonumber \\
\label{ward}
\end{eqnarray}
where summations are not taken over the repeated indices on the
r.h.s.

\hspace*{1ex}

\noindent $\bullet$ {\em Thermal self-energy part of a quark and
resummed quark propagator}

The quasiparticle or diagonalized self-energy part
$\tilde{\Sigma}_F$ is related \cite{lan} to the so-called analytic
self-energy part $\Sigma$ through $\tilde{\Sigma}_F (r_0, r) =
\Sigma (r_0 (1 + i 0^+), r)$. In real-time thermal field theory,
$\tilde{\Sigma}_F$ is evaluated \cite{lan} through $\tilde{\Sigma}_F
(r_0, r) = \Sigma_{1 1} (r_0 (1 + i 0^+), r)$ $+$ $\Sigma_{1 2}
(r_0 (1 + i 0^+), r)$. Decomposing $\tilde{\Sigma}_F$ as
\begin{equation}
\tilde{\Sigma}_F (r_0, r) = a (r_0, r)
\Rslash + b (r_0, r) \, \gamma^0 \, , \label{sigma}
\end{equation}
we obtain, to one-loop order,
\begin{eqnarray}
a (r_0, r) & = & - \, \frac{m_f^2}{r^2} \left( 1 - \frac{r_0}{2 r}
\, \ln \frac{r_0 + r}{r_0 - r} \right) + \tilde{a} (r_0, r) \, ,
\nonumber \\
b (r_0, r) & = & m_f^2 \, \frac{r_0}{r^2} \left( 1 -
\frac{R^2}{2 r_0 r} \, \ln \frac{r_0 + r}{r_0 - r} \right) + R^2 \,
\tilde{b} (r_0, r) \, ,
\nonumber \\
\tilde{a} (r_0, r) & = & g^2 \, \frac{N_c^2 - 1}{2 \, N_c} \,
\frac{1}{r^2}
\left[ (r_0^2 + r^2) I_B + R^2 I_F - 2 r_0 (J_B + J_F)
\right] \, , \label{a-tilde} \\
\tilde{b} (r_0, r) & = & - g^2 \, \frac{N_c^2 - 1}{2 \, N_c} \,
\frac{1}{r^2}
\left[ r_0 (I_B + I_F) - 2 (J_B + J_F)
\right] \, ,
\label{b-tilde}
\end{eqnarray}
with
\begin{eqnarray}
I_B & = & \frac{1}{16 \pi^2} \, \frac{1}{r} \int_0^\infty dk \,
n_B (k) \ln \left\{ \prod_{\xi = \pm}
\frac{r_0 + \xi (r - 2 k)}{r_0 - \xi (r + 2 k)} \right\} \, ,
\label{inti} \\
J_B & = & \frac{1}{16 \pi^2} \, \frac{1}{r} \int_0^\infty dk \, k \,
n_B (k) \ln \left\{ \prod_{\xi = \pm} \frac{r_0 + r - 2
\xi k}{r_0 - r - 2 \xi k} \right\} \, .
\label{int}
\end{eqnarray}
$I_F$ and $J_F$ in (\ref{a-tilde}) and (\ref{b-tilde})are given,
respectively, by $I_B$ and $J_B$ with $n_B ( k ) \to n_F ( k )$. It
is to be noted that, when $R$ is soft, $I_B$, $J_B$, $I_F$ and $J_F$
may be neglected in the HTL approximation, and we can reproduce
(\ref{eff1}) - (\ref{eff9}).

The self-energy-part corrected propagator of a quark
$\displaystyle{ \raisebox{1.1ex}{\scriptsize{$\diamond$}}}
\mbox{\hspace{-0.33ex}} S_{j \ell} (R)$ is written \cite{lan} as in
(\ref{eff1}) - (\ref{eff9}), provided $D_\sigma$ in (\ref{eff5}) is
replaced by
\begin{eqnarray}
D_\sigma (R) & \rightarrow & \left( - r_0 + \sigma r \right)
\left\{ 1 - a (r_0, r) \right\} + b (r_0, r) \, \nonumber \\
& & \mbox{\hspace{2ex}} = D_\sigma (R) + (r_0 - \sigma r) \,
\tilde{a} + R^2 \,
\tilde{b} \, .
\label{hardeff}
\end{eqnarray}


\newpage
\begin{center}
{\Large {\bf Figure captions} } \vspace*{1.5em} \\
\end{center}

\begin{description}
\item[Fig. 1.] Effective one-loop diagram for the production of a
real soft photon. The blobs indicate the effective quark propagators
and the effective quark-photon vertices. $1$ and $2$ at the
effective photon-quark vertices designate the type of vertex.
\item[Fig. 2.] Structure of the HTL correction,
$\displaystyle{\raisebox{0.6ex}{\scriptsize{*}}}
\tilde{\Gamma}^\mu_{2 1} (K', K)$, to the photon-quark vertex.
\item[Fig. 3.] An example of the physical processes taking place
in a quark-gluon plasma, which is responsible for the singular
contribution (\ref{w2}) to $E \, dW / d^{\, 3} p$. The left side of
the final-state cut line, $C_1$ or $C_2$, represents the $S$-matrix
element, while the right side represents the $S^*$-matrix element.
For the diagram with the final-state cut line $C_1$, when the
effective quark propagator
$\displaystyle{ \raisebox{1.1ex}{\scriptsize{$\diamond$}}}
\mbox{\hspace{-0.33ex}} S_{11} (Q + P)$ (cf. (\ref{S-star1})) is
used, the mass singularity is screened. As to the photons and gluons
in the diagram, in addition to the transverse components, the
longitudinal and scalar components are included, because of the
factor $g_{\mu \nu}$ in (\ref{def}) and of the usage of the Feynman
gauge. If we use the Coulomb gauge and $g^{(t)}_{\mu \nu}$ (eq.
(\ref{g0})) in place of $g_{\mu \nu}$ in (\ref{def}), only
transverse components take place. In this case, the diagrams which
are topologically different from Fig. 1 (and then also this
diagram) contribute to the singular contributions.
\end{description}

\end{document}